\documentclass[aps,prb,twocolumn,superscriptaddress]{revtex4-2}

\usepackage[bookmarks=false,colorlinks=true,urlcolor=blue,citecolor=blue,linkcolor=blue]{hyperref}

\usepackage{cancel}
\usepackage{xspace}
\usepackage{tikz}
\usetikzlibrary{decorations.pathmorphing,patterns}
\usetikzlibrary{calc,arrows}
\usepackage[compat=1.1.0]{tikz-feynman}
\tikzfeynmanset{/tikzfeynman/warn luatex=false}

\usepackage{amsmath,amssymb,bm}
\usepackage{gensymb}
\usepackage[normalem]{ulem}
\usepackage{booktabs}

\usepackage[sectionbib]{bibunits}
\defaultbibliographystyle{apsrev4-2}
\defaultbibliography{kag}

\usepackage{graphicx}

\newcommand{\mycomment}[1]{}

\newcommand{\be}{\begin{equation}}
\newcommand{\ee}{\end{equation}}
\newcommand{\bea}{\begin{eqnarray}}
\newcommand{\eea}{\end{eqnarray}}
\newcommand{\bes}{\begin{equation} \begin{split}}
\newcommand{\ees}{\end{split} \end{equation}}

\newcommand{\f}{\frac}

\newcommand{\qv}{\vec{q}}
\newcommand{\qvp}{\vec{q}^{\,\prime}}

\newcommand{\Qv}{\vec{Q}}
\newcommand{\pv}{\vec{p}}

\newcommand{\rv}{\vec{r}}

\newcommand{\Rv}{\vec{R}}
\newcommand{\Rvp}{\vec{R}^{\prime}}

\newcommand{\Sv}{\vec{S}}

\newcommand{\av}{\vec{a}}

\newcommand{\latentheat}{\ell_h}

\makeatletter
\newcommand{\Kinv}{\@ifnextchar\bgroup{\Kinvargs}{K^{-1}}}
\newcommand{\Kinvargs}[3]{K^{-1 #1 #2}_{#3}}
\newcommand{\Dinv}{\@ifnextchar\bgroup{\Dinvargs}{D^{-1}}}
\newcommand{\Dinvargs}[3]{D^{-1 #1 #2}_{#3}}

\newcommand{\Kbare}{\mathcal{K}}
\newcommand{\Kbaremat}{\bm{\Kbare}}
\newcommand{\Kbareinv}{\Kbare^{-1}}
\newcommand{\Kbareinvmat}{\bm{\Kbareinv}}
\newcommand{\Lambdabare}{\Lambda}
\newcommand{\Lambdabaremat}{\bm{\Lambdabare}}
\newcommand{\Dbare}{\mathcal{D}}

\newcommand{\Dbareinv}{\Dbare^{-1}}
\newcommand{\Dbareinvmat}{\bm{\Dbareinv}}

\newcommand{\Trace}{\mathrm{Tr}}
\newcommand{\trace}{\mathrm{tr}}

\newcommand{\rthree}{${\sqrt{3} \! \times \! \! \sqrt{3}}$\xspace}

\newcommand{\RN}[1]{%
\textup{\uppercase\expandafter{\romannumeral#1}}%
}

\newcommand{\cv}{c_{\rm v}}

\usepackage{contour}

\definecolor{taylorswift}{rgb}{0.0862745098,0.4666666667,0.3411764706}
\definecolor{fearless}{rgb}{0.8862745098,0.6117647059,0.2823529412}
\definecolor{speaknow}{rgb}{0.4588235294,0.2274509804,0.4980392157}
\definecolor{red}{rgb}{0.6509803922,0.1254901961,0.2705882353}
\definecolor{TS1989}{rgb}{0.1803921569,0.6,0.9764705882}
\definecolor{reputation}{rgb}{0.1450980392,0.1490196078,0.1529411765}
\definecolor{lover}{rgb}{0.8392156863,0.2117647059,0.5529411765}

\tikzset{>=latex}

\begin{document}

\title{Weak first-order phase transition out of the classical kagome spin liquid}

\author{Cecilie Glittum}
\affiliation{Department of Physics, University of Oslo, P.~O.~Box 1048 Blindern, N-0316 Oslo, Norway}
\author{Olav F. Sylju{\aa}sen}
\affiliation{Department of Physics, University of Oslo, P.~O.~Box 1048 Blindern, N-0316 Oslo, Norway}

\date{\today}

\begin{abstract}
The low-temperature fate of the spin-liquid regime in the classical kagome Heisenberg antiferromagnet has been debated for over three decades. Using an expansion in the number of spin components, we show that, contrary to earlier Monte Carlo simulations, the spin liquid terminates at a weak first-order phase transition into the \rthree phase which ordered moment saturates at zero temperature. Adding second-neighbor interactions, this transition belongs to a line of first-order phase transitions that ends at a critical point. For comparison, the pyrochlore antiferromagnet remains disordered at all temperatures.
\end{abstract}

\maketitle

\begin{bibunit}


Magnetic frustration arises when the geometry of a crystal lattice prevents spins from simultaneously minimizing all their pairwise interactions. Rather than freezing into a conventional ordered state, frustrated magnets can remain in a highly correlated yet disordered cooperative regime, a classical spin liquid, down to temperatures far below the energy scale set by the exchange interactions. A central question is whether such a spin liquid is ultimately stable all the way down to zero temperature, or whether thermal fluctuations eventually conspire to select an ordered state at anomalously low temperatures~\cite{Lacroix2011}.

\begin{figure}[]
\centering
\includegraphics[width=0.85\columnwidth]{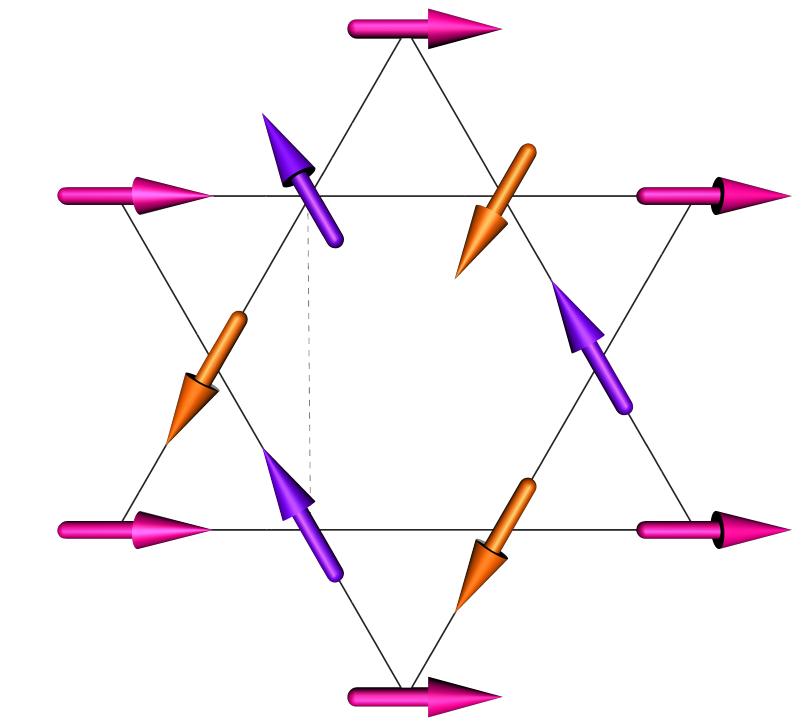}
\caption{Kagome lattice with spins showing the coplanar fully ordered \rthree state. Solid lines show $J_1$ couplings. The dashed line shows one of the $J_2$ couplings.}
\label{fig:kagomelattice}
\end{figure}

The classical Heisenberg antiferromagnet on the kagome lattice, see Fig.~\ref{fig:kagomelattice}, is the sharpest two-dimensional realization of this problem. Its corner-sharing triangle geometry forces every triangle to have zero total spin, leaving a ground-state manifold that is extensively degenerate. The resulting spin liquid, characterized by a diffuse neutron-scattering intensity, persists over a broad temperature range well below the exchange interaction scale. Yet both analytical arguments and numerical Monte Carlo (MC) simulations find that, at sufficiently low temperatures, thermal fluctuations select coplanar states and ultimately favor \rthree correlations~\cite{Harris1992,Chalker1992,Huse1992,Sachdev1992,Reimers1993,Zhitomirsky2008,Henley2009,Chern2013}, see Fig.~\ref{fig:kagomelattice}, through an order-by-disorder mechanism~\cite{Villain1980}.

Despite three decades of sustained efforts, two fundamental questions about the low-temperature behavior have remained delicate to settle: First, does the spin liquid evolve into the \rthree phase through a thermal crossover, or is there a thermodynamic phase transition separating two distinct phases? Second, does the \rthree ordered moment truly saturate as the temperature goes to zero, or does it remain permanently suppressed by proliferating domain walls~\cite{Reimers1993} and/or vortices~\cite{Zhitomirsky2008}?
Extensive MC efforts have led to the prevailing picture for both questions: the system appears to undergo a sharp crossover into a coplanar phase with a saturating octupolar order parameter~\cite{Ritchey1993,Zhitomirsky2008} and \rthree correlations with a substantially reduced ordered moment~\cite{Reimers1993,Zhitomirsky2008,Chern2013}.
However, as noted in Refs.~\onlinecite{Reimers1993,Zhitomirsky2008}, such MC simulations are challenging, as the MC algorithms experience severe slowing down at low temperatures. This motivates the use of complementary methods that do not rely on direct configuration sampling.

In this Letter, we address these long-standing problems using Nematic Bond Theory (NBT), a self-consistent approach based on the large-$N_s$ expansion where $N_s=3$ is the number of spin components.
It extends the usual self-consistent Gaussian approximation (SCGA)~\cite{Chalker2017} by including momentum-dependent self-energies and enforcing the local unit-length spin constraint more accurately~\cite{Schecter2017,Syljuasen2019,Glittum2021,Glittum2023}. NBT is particularly well suited to the kagome problem because it compares free energies of competing phases directly for large system sizes, thus bypassing sampling problems that can impair low-temperature MC studies. Our main result is that the classical kagome spin liquid and the \rthree phase are genuine thermodynamic phases separated by a first-order transition with very small latent heat. Furthermore, the ordered moment of the \rthree phase saturates as the temperature goes to zero.

We consider the classical Heisenberg model
\begin{equation}
H=\frac12\sum_{\Rv,i}\sum_{\Rvp,j}J_{\Rvp-\Rv, ij}\,\Sv_{\Rv,i}\cdot\Sv_{\Rvp,j},
\end{equation}
on the kagome lattice with unit-length spins $\Sv_{\Rv,i}$, where $\Rv$ denotes the unit cell and $i$ the position within the unit cell (sublattice index).
The lattice has $N=3L^2$ spins and periodic boundary conditions. We focus on nearest-neighbor antiferromagnetic coupling $J_1=1$, with a second-neighbor coupling $J_2$ introduced later to map the nearby phase diagram. While SCGA takes the unit-length constraint of the spins into account approximately by enforcing unit length on average, NBT goes one step further by also suppressing spatial fluctuations of the spin lengths. To do this, the local spin-length constraint is represented by a fluctuating constraint field; its uniform component $\Delta$ is treated self-consistently as in SCGA, while nonuniform components are incorporated diagrammatically. This yields dressed spin and constraint propagators linked by self-consistent Dyson equations with a momentum-dependent self-energy (see Supplemental Material Sec.~A~\cite{suppmat}).
Solving these NBT equations amounts to finding a saddle-point solution by means of iterations, starting from a chosen value of the self-consistent field $\Delta$ and an initial self-energy. The converged solution in turn gives the temperature, spin static structure factor and crucially the free energy. Operationally, we iterate the NBT equations from different initial self-energies and track the converged branches as $\Delta$ is varied. Because the iterations are not guaranteed to flow to the globally stable state, a key step is to compare the resulting free energies, and identify the physical branch as the one with the lowest value.

\begin{figure}[]
\begin{center}
\includegraphics[width=\columnwidth]{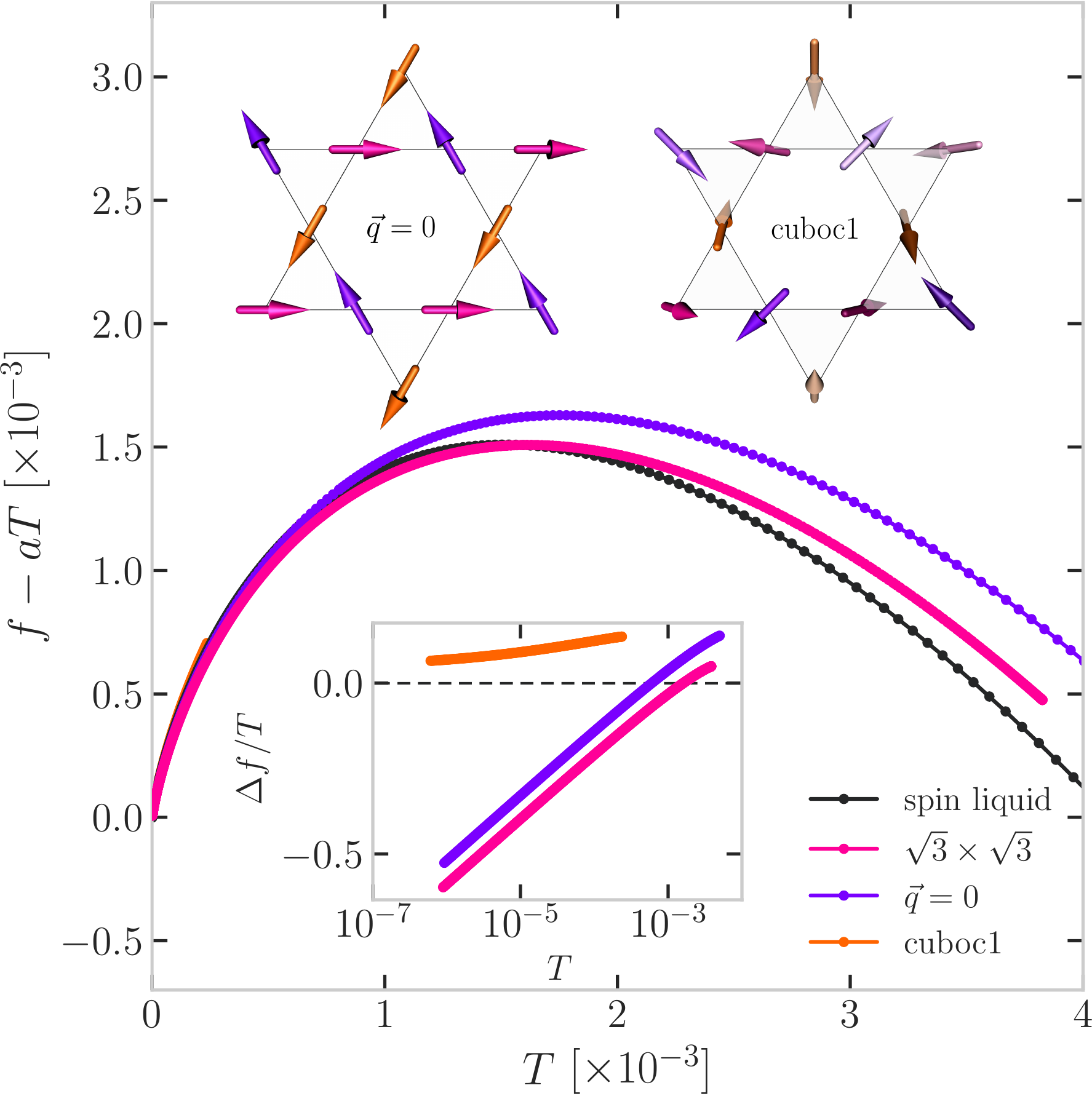}
\caption{Main panel: Free energy per spin vs. $T$ for the different phases indicated by the legends. The black points start from a random initial self-energy at large values of $\Delta$, and constitute the free energy of the spin liquid. The colored points are obtained starting from low $\Delta$-values with initial self-energies that bias the iterations toward their respective phases.
A linear term $a T$ with $a=3$ is subtracted for better visualization of the crossing point between the spin-liquid and \rthree branches.
  The inset shows the difference of the biased curves from the spin liquid divided by $T$. $L=300$. The spin configurations show the coplanar $\qv=0$ state (left) and the noncoplanar cuboc1 state (right). 
}
\label{fig:freeenergy}
\end{center}
\end{figure}

\begin{figure}[]
\begin{center}
\includegraphics[width=\columnwidth]{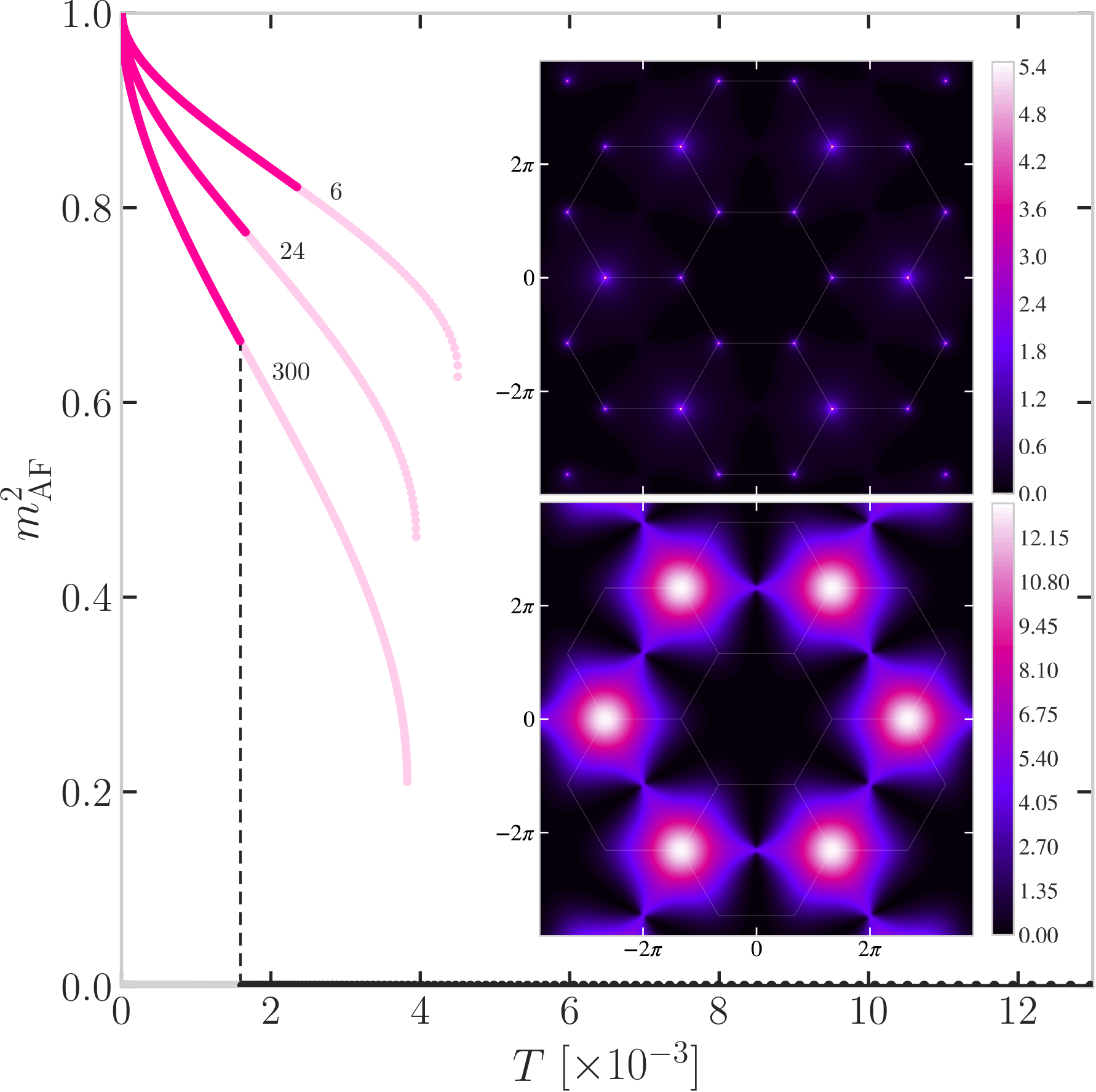}
\caption{Main panel: The (squared) ordered moment for the \rthree phase (pink) and spin liquid (black) vs. $T$ for different linear system sizes $L$ indicated by the numbers. The vertical line marks $T_c$ for $L=300$. Faint colors correspond to metastable states. Insets: Spin structure factors $S(\qv)$ for the \rthree phase at $T=1.58 \cdot 10^{-3}$ (top, $\log_{10}(S(\qv)+1)$), and for the spin liquid at $T=1.78 \cdot 10^{-3}$ (bottom, $S(\qv)$). $L=300$.}
\label{fig:orderpar}
\end{center}
\end{figure}

For large $\Delta$, which corresponds to high temperatures $T$, the NBT equations converge to a unique solution regardless of the initial self-energy. The corresponding free energy is shown as the black curve in Fig.~\ref{fig:freeenergy}. For points on this curve, our NBT results are approximately equal to SCGA and describe a disordered paramagnet at high temperatures that evolves smoothly into a highly fluctuating, noncoplanar spin liquid with a diffuse structure factor and clear pinch points at lower temperatures, see Fig.~\ref{fig:orderpar} lower inset~\cite{Sachdev1992,Garanin1999,Zhitomirsky2008}. This spin-liquid branch persists down to the lowest temperatures in our calculation; however, its free energy is minimal only at temperatures $T>T_c \approx 1.59 \cdot 10^{-3}$. At lower temperatures, the coplanar single-$\qv$ \rthree phase becomes thermodynamically stable, see the pink curve in Fig.~\ref{fig:freeenergy}. This curve is obtained by starting from a low value of $\Delta$ and an initial self-energy that biases the iterations toward the \rthree state. To find this initial self-energy, we performed a simulation with $J_2$ slightly ferromagnetic, and used the converged self-energy at low temperature as input.
States in the \rthree phase have structure factors with sharp peaks at the \rthree ordering wave vectors and corresponding satellite peaks, see Fig.~\ref{fig:orderpar} upper inset. The associated correlation length in the \rthree phase at the phase transition is very large, exceeding $3000$ kagome lattice spacings. In contrast, the correlation length in the spin liquid at the phase transition is approximately twenty kagome lattice spacings and follows a $1/\sqrt{T}$ behavior~\cite{Garanin1999} (see Supplemental Material Sec.~B~\cite{suppmat}).

The fact that the free energies of the spin liquid and \rthree phase cross at a very low, but finite, temperature, implies that they are distinct thermodynamic phases. The discontinuity of slopes at the crossing point further indicates a first-order phase transition.
We have carried out a finite-size analysis of the crossing point (see Supplemental Material Sec.~C~\cite{suppmat}), and find that the first-order transition has a very small latent heat per spin $\latentheat = 1.04 \cdot 10^{-4}$, thus classifying the transition as {\em weakly} first order.
 
Figure~\ref{fig:freeenergy} shows also two other free-energy branches: the purple and orange curves where the initial self-energy biases the iterations toward the $\qv=0$ state and the cuboc1 state~\cite{Janson2008,Messio2011}, respectively (for illustrations, see top of Fig.~\ref{fig:freeenergy}). Their free energies are always larger than the \rthree phase, which is even more apparent in the lower inset of Fig.~\ref{fig:freeenergy} showing a semilog-plot of the difference in free energy from the spin liquid divided by temperature.

To further characterize the low-temperature \rthree phase, we calculate the (same sublattice) \rthree (dipolar) ordered moment squared~\cite{Zhitomirsky2008}
\be
m^2_{\rm AF} \equiv \f{6}{N^2} \sum_{l,\Rv,\Rvp} \langle \vec{S}_{\Rv,l} \cdot \vec{S}_{\Rvp,l} \rangle e^{i \Qv \cdot \left( \Rv - \Rvp \right)},
\ee
where $\Qv=(4\pi/3,0)$ corresponds to antiferromagnetic \rthree order.
Fig.~\ref{fig:orderpar} shows $m^2_{\rm AF}$ as function of temperature for three different system sizes. As the system is two-dimensional with continuous symmetry, the ordered moment must vanish at finite temperatures as $N \to \infty$~\cite{Mermin1966}. However as is also apparent from the figure, it saturates at $T=0$ for all system sizes, i.e. the \rthree phase orders completely at $T=0$. In contrast, $m^2_{\rm AF}$ calculated in the spin lquid is almost zero.

\begin{figure}[]
\begin{center}
\includegraphics[width=0.9845\columnwidth]{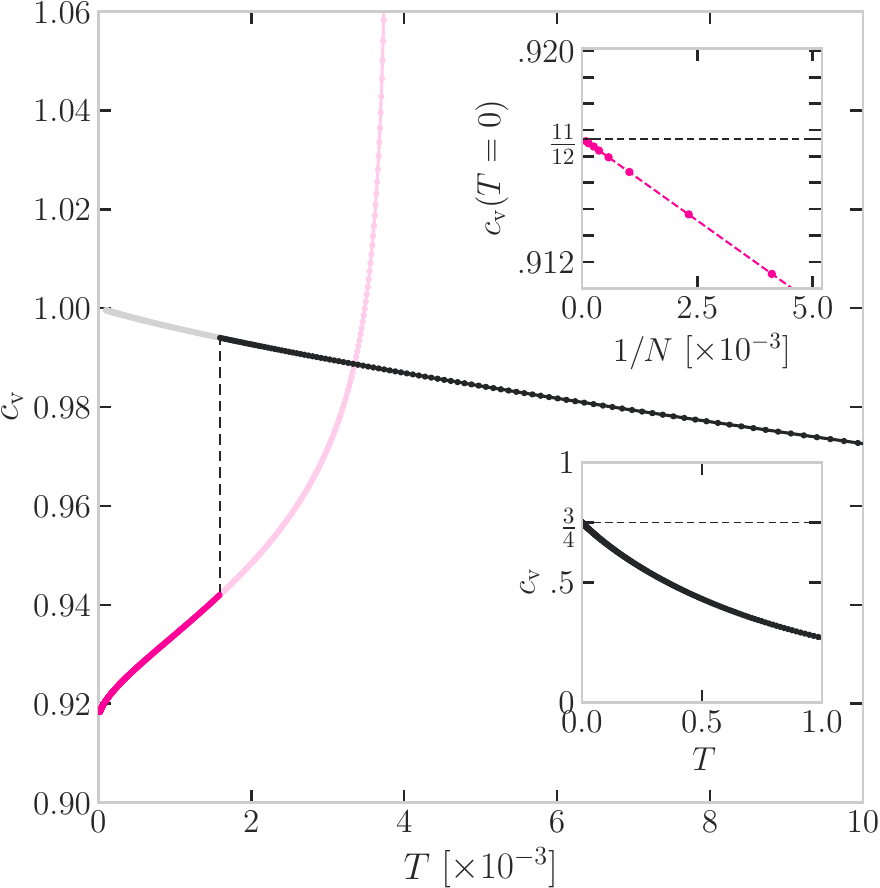}
\caption{Main panel: The specific heat vs. $T$ for a system with linear size $L=300$. Black points are derived from the spin-liquid branch in Fig.~\ref{fig:freeenergy} while pink points come from the corresponding \rthree branch. Fainter colors mark when a given branch becomes metastable.
Upper inset: Specific heat at $T=0$ for different inverse system sizes. Lower inset: Specific heat vs. $T$ for the pyrochlore lattice antiferromagnet with $4 \times 36^3$ sites.}
\label{fig:specificheat}
\end{center}
\end{figure}

Next we obtain the specific heat $\cv$ as a function of temperature from the free energy, focusing on the thermodynamically stable branches, see Fig.~\ref{fig:specificheat}. When lowering the temperature, the specific heat of the spin liquid increases and extrapolates to $1$ at $T=0$. At $T_c$ the first-order phase transition manifests itself in $\cv (T)$ as a Dirac delta-function $\latentheat \delta(T-T_c)$ (not shown) and a discontinuity (vertical dashed line). Below $T_c$ the specific heat follows the \rthree branch and decreases steadily. For temperatures close to zero, the specific heat follows a $\sqrt{T}$ behavior (see Supplemental Material Sec.~D~\cite{suppmat}). The zero-temperature values for different system sizes are shown in the upper inset and a straight line fit to them gives $0.9167 - 1.2495/N$. This agrees well with the predicted behavior $11/12 + 5/4N$ for coplanar states~\cite{Chalker1992}. The apparent divergence of the metastable \rthree specific heat indicates a possible pseudo-critical point. No such divergence in $\cv$ is present for the metastable spin liquid.

A comparison between NBT and MC reveals two central differences: First, the low-temperature state found by NBT is quantitatively different from the one inferred from earlier MC simulations. Despite both approaches showing \rthree correlations, the ordered moment saturates within NBT, whereas in MC simulations it remains strongly suppressed at about 10\% of the NBT values~\cite{Zhitomirsky2008,Chern2013}.
Second, instead of showing a clear signature of a weak phase transition, MC simulations of $\cv$ show a broad plateau that ends around $T \approx 4\cdot 10^{-3}$, followed by a steep decrease~\cite{Zhitomirsky2008,Schnabel2012}.

In MC simulations, a latent heat delta-function would appear as a peak at $T_c$ with height $(\ell_{h}/2T_c)^2 N \approx 10^{-3} N$ and width $(2T_c)^2/(\ell_{h} N )  \approx 0.1/N$~\cite{Imry1980,Janke2003}. 
Such a peak is hard to capture in finite-size MC simulations as it will be very narrow and sit on top of a discontinuity. Furthermore, MC equilibration across different domain-wall sectors is presumably slow at low temperatures. We therefore consider it plausible that, in the vicinity of the phase transition, conventional MC simulations may not be fully equilibrated and that the resulting configurations contain domains of both phases. This picture of phase coexistence near a weak first-order transition is consistent with the structure factor observed in MC simulations, which is reminiscent of a mixture of the \rthree structure factor and the intermediate-temperature spin-liquid structure factor~\cite{Zhitomirsky2008}. It should nevertheless be mentioned that the onset of coplanar ordering in MC simulations has been suggested to correspond to a topological (phase) transition~\cite{Zhitomirsky2008}.

The permanently reduced ordered moment in MC simulations is also consistent with phase coexistence, but has been attributed instead to the proliferation of coplanar domain walls between \rthree domains with opposite chirality, while keeping the system within the ground-state manifold~\cite{Reimers1993}, consistent with a saturating octupolar order parameter as $T \to 0$. However, it is not clear that domain-walls will proliferate at low temperatures in equilibrium. Even though domain-wall states have a large configurational multiplicity and therefore substantial entropic weight, domain walls stiffen the spin configuration and thereby reduce the entropic advantage compared to the \rthree state~\cite{Henley2009}. This subtle competition between the configurational entropy of domain-wall states and the entropic gain of soft modes in the \rthree state is difficult to settle.

\begin{figure}[]
\begin{center}
\includegraphics[width=\columnwidth]{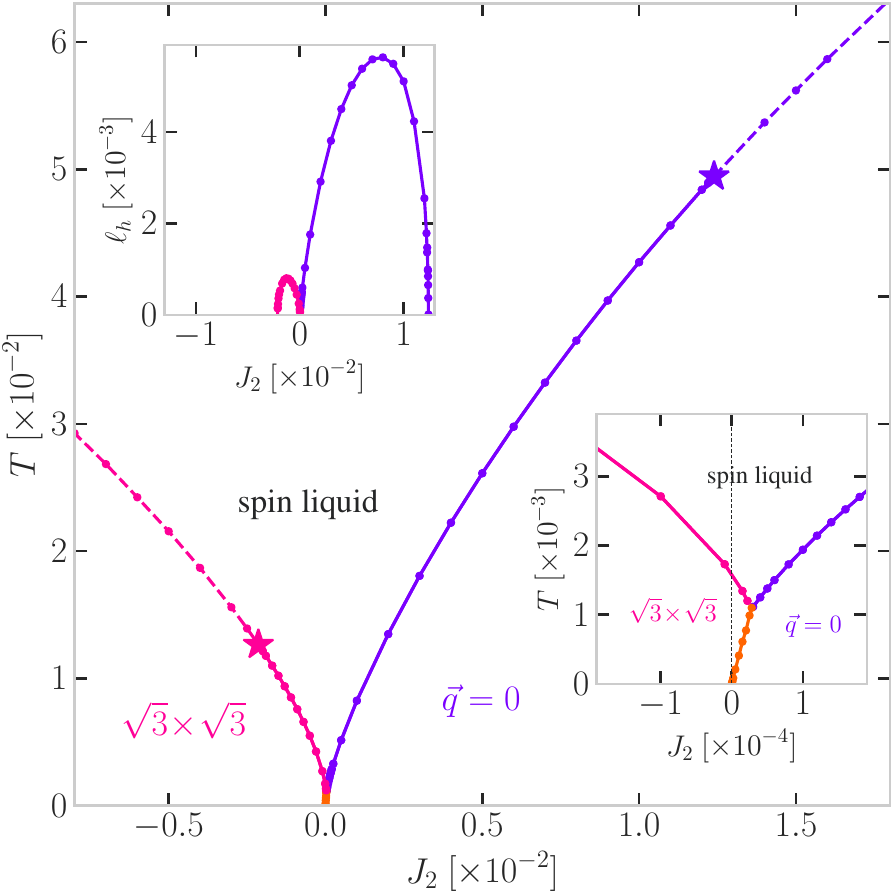}
\caption{Main panel: $J_2$-$T$ phase diagram. Solid lines indicate weak first-order phase transitions and dashed lines indicate crossovers. Critical points are marked by stars. Lower right inset: Blow-up of the region around the triple point. Upper left inset: Latent heat per spin of the phase transitions from the spin liquid to the ordered phases as a function of $J_2$.}
\label{fig:phasediagram}
\end{center}
\end{figure}

To test the robustness of the weak first-order phase transition, we extend our analysis to finite second-neighbor coupling $J_2$. The resulting $J_2$-$T$ phase diagram in Fig.~\ref{fig:phasediagram} contains two finite lines of first-order transitions that separate the disordered spin-liquid regime from two distinct low-temperature ordered phases: the \rthree phase for ferromagnetic $J_2 < 0$ and the $\qv=0$ phase for antiferromagnetic $J_2 > 0$. A blow-up of the small $J_2$-region, see right inset of Fig.~\ref{fig:phasediagram}, further reveals that the \rthree phase extends slightly into the antiferromagnetic $J_2$ side at finite temperatures, consistent with the \rthree state having higher spin-wave entropy than the $\qv=0$ state~\cite{Henley2009}. The fact that the \rthree phase protrudes slightly into the region $J_2 > 0$ naturally explains the existence of the first-order transition in the nearest-neighbor model ($J_2 = 0$). The two first-order lines meet in a triple point at $(J_2,T_c)=(2.8\cdot10^{-5},1.11\cdot10^{-3})$ where the spin liquid, \rthree and $\qv=0$ phases coexist with a finite, but small $\sim 10^{-4}$, latent heat. The \rthree and $\qv=0$ phases are separated by a third line of first-order transitions.

The upper inset of Fig.~\ref{fig:phasediagram} shows the latent heat per spin associated with the spin liquid to (local) order transitions. They are all weakly first order ($\latentheat < 6 \cdot 10^{-3}$), and exist only over a limited range of $J_2$ before they terminate at critical points, marked by stars at
$(J_2,T_c)=(-2.139\cdot 10^{-3},1.27\cdot10^{-2})$ and $(1.239\cdot 10^{-2},4.94\cdot 10^{-2})$,
where the latent heat vanishes. Outside of the critical points we do not detect phase transitions, but instead crossovers associated with a finite peak in the specific heat that diverges at the critical points.

This extension partly agrees with earlier MC work on the $J_1$-$J_2$ kagome model~\cite{Spenke2018}. For antiferromagnetic couplings $J_2 < 7.5\cdot 10^{-3}$, MC simulations and NBT both find first-order phase transitions into the $\qv=0$ phase, with comparable latent heats and transition temperatures. In particular, at $J_2 = 5 \cdot 10^{-3}$, the NBT values of $\latentheat = 5.02 \cdot 10^{-3}$ and $T_c = 2.61\cdot 10^{-2}$ are close to the MC values. However, for large $J_2$, both ferromagnetic and antiferromagnetic, NBT indicates only crossovers, whereas MC simulations have been interpreted in terms of 2D Ising–type continuous phase transitions. Our results suggest that the apparent scaling indicative of such behavior is instead a finite-size effect that disappears for sufficiently large system sizes (see Supplemental Material Sec.~E~\cite{suppmat}). For small ferromagnetic $J_2$, the MC simulations report a double-peak structure in the specific heat as a function of temperature. The high-temperature peak has been associated with a crossover into a \rthree phase of reduced ordered moment, while there is an additional unclassified transition at lower temperature into the fully ordered \rthree phase. NBT, by contrast, finds a single first-order phase transition into the \rthree phase. 

For comparison, we have applied NBT also to the nearest-neighbor pyrochlore Heisenberg antiferromagnet, the canonical three-dimensional classical spin liquid~\cite{Villain1979, Reimers1991, Reimers1992, Moessner1998PRL}. In that case we find no analogous ordered branch that overtakes the spin liquid at low temperatures (see Supplemental Material Sec.~F~\cite{suppmat}). Within NBT, the pyrochlore antiferromagnet remains disordered down to the lowest temperatures studied, with a specific heat approaching the value $0.7500 - 1.5001/N$, see lower inset of Fig.~\ref{fig:specificheat}. The infinite-size value agrees well with $3/4$ reported in Ref.~\onlinecite{Moessner1998PRL}. This contrast emphasizes that the kagome result is not an artifact of the NBT method generically producing spurious order in highly frustrated systems; instead, it reflects a genuine distinction between two archetypal classical spin liquids.

It is appropriate to mention that the NBT method is approximate, as the diagrammatic treatment of the constraint field excludes vertex corrections (see Supplemental Material Sec.~A~\cite{suppmat}).
Nevertheless, NBT has previously produced the correct phases and phase transitions for other frustrated models~\cite{Glittum2023, Glittum2026}, but typically overestimates critical temperatures by 10-20\%~\cite{Syljuasen2019}. Thus, all temperatures presented in this Letter should be understood as estimates.
As NBT relies on convergence from an initial self-energy, and different initial self-energies could realize different converged phases, there is always a possibility that we have not found the correct thermodynamically stable state due to bad initialization. We have therefore also performed NBT simulations with initial self-energies obtained from MC simulations. However, instead of stabilizing an alternative phase, they all converge to either the spin liquid or to the \rthree phase (see Supplemental Material Sec.~G~\cite{suppmat}).

In conclusion, we have used NBT to shine new light on a critical problem in classical frustrated magnetism: the nearest-neighbor kagome Heisenberg antiferromagnet does not merely drift into a coplanar regime through crossover physics, but exits its spin-liquid regime through a genuine, though weak, first-order phase transition. The low-temperature phase is the \rthree phase which ordered moment saturates as $T\to 0$.\\

C.G. acknowledges funding from the European Union’s Horizon Europe research and innovation programme under the Marie Skłodowska-Curie Grant Agreement No. 101126636.
The computations were performed on resources provided by Sigma2 - the National Infrastructure for High Performance Computing and Data Storage in Norway, and on the Fox supercomputer at the University of Oslo.


\putbib
\end{bibunit}

\clearpage

\onecolumngrid

\begin{bibunit}

\section*{Supplemental material}

\subsection{Nematic Bond Theory}
The Nematic Bond Theory (NBT)~\cite{Schecter2017,Syljuasen2019,Glittum2021,Glittum2023} is an extension of the self-consistent Gaussian approximation (SCGA)~\cite{Chalker2017} that goes beyond it by suppressing fluctuations in the lengths of the spins. To explain how it is used in this paper we begin by formulating it on the kagome lattice.

The kagome lattice can be treated as a triangular Bravais lattice with a three-site unit cell. The coordinate of the unit cell will be denoted $\Rv$ and can be expressed in terms of the triangular Bravais lattice vectors $\vec{a}_1 = (1,0)$ and $\vec{a}_2 = (-1/2,\sqrt{3}/2)$. For convenience, we also define $\vec{a}_3= - \av_1 - \av_2 = (-1/2,-\sqrt{3}/2)$. We have chosen the lattice spacing on the Bravais lattice to be unity, which means that the lattice spacing on the kagome lattice itself is one half. A site within a unit cell will be indexed by its sublattice index $i \in \{1,2,3\}$ so that its position within the unit cell $\vec{\alpha}_i$ takes one of the three values $\{ (0,0), (1/2,0), (1/4,\sqrt{3}/4) \}$. A site on the kagome lattice can then be uniquely defined by its sublattice index $i$ and unit cell $\Rv$, so that for the spin at $\rv = \Rv + \vec{\alpha}_i$, we write $\Sv_{\rv = \Rv+ \alpha_i} \equiv \Sv_{\Rv,i}$.

The Hamiltonian of the Heisenberg model is
\be
H = \f{1}{2} \sum_{\Rv,i} \sum_{\Rvp,j}  J_{\Rvp-\Rv, ij} \Sv_{\Rv,i} \cdot \Sv_{\Rvp,j}, \label{supp:HeisenbergHamiltonian}
\ee
where $J_{\Rvp-\Rv, ij}$ denotes the exchange coupling between the spin at sublattice $i$ in unit cell $\Rv$ and the spin at sublattice $j$ in unit cell $\Rvp$. The factor $1/2$ is inserted to avoid double counting. 
We then introduce Fourier transforms
\begin{align}
\Sv_{\Rv,i} &= \f{1}{\sqrt{N_c}} \sum_{\qv} \Sv_{\qv,i} e^{i \qv \cdot \Rv}, \\
J_{\Rv,ij} &= \f{2}{N_c} \sum_{\qv} J_{\qv,ij} e^{-i \qv \cdot \Rv},
\end{align}
where the $\qv$-sums go over the first Brillouin zone of the triangular Bravais lattice, and $N_c$ is the total number of unit cells so that the total number of spins is $N = 3 N_c$.
The Fourier conventions have been chosen differently for $\Sv$ and $J$ for notational convenience. 
The Hamiltonian can then be written compactly
\be
H = \sum_{\qv} \sum_{ij} J_{\qv,ij} \Sv^*_{\qv,i} \cdot \Sv_{\qv,j}, \label{supp:qHamiltonian} 
\ee
where $J_{\qv,ij}$ can be interpreted as the components of a three-by-three matrix with $\qv$-dependent entries, and we have used $\Sv_{-\qv,i} = \Sv^*_{\qv,i}$. For the kagome lattice the matrix $2J_{\qv}$ is
\be
\begin{pmatrix}
  0 & J_1(1+e^{-i q_1}) +J_2 (e^{i q_2} + e^{i q_3}) & J_1 (1+e^{i q_3}) +J_2 (e^{-i q_1} + e^{-i q_2}) \\
  J_1(1+e^{i q_1}) +J_2 (e^{-i q_2} + e^{-i q_3}) & 0 & J_1 (1+e^{-i q_2}) +J_2 (e^{i q_3} + e^{i q_1}) \\
  J_1 (1+e^{-i q_3}) +J_2 (e^{i q_1} + e^{i q_2}) & J_1 (1+e^{i q_2}) +J_2 (e^{-i q_3} + e^{-i q_1})& 0
\end{pmatrix},
\ee
where $q_i = \qv \cdot \vec{a}_i$. In order to ensure that the matrix $J_{\qv}$ is positive definite we will redefine it by subtracting its minimum eigenvalue. This corresponds to subtrating a constant in the Hamiltonian, and has no consequence for the physical properties of the model.  

To obtain the free energy we will compute the partition function for the canonical ensemble $Z$ by integrating over all spins
\be
   Z = \int \prod_{\Rv,i} d\Sv_{\Rv,i} e^{-\beta H}.
\ee
The unit-length constraint on the spins will be taken into account by delta-functions written as an integral over a constraint field $\lambda_{\Rv,i}$ at each site:
\begin{align}
\prod_{\Rv,i} \delta( | \Sv_{\Rv,i} | -1 ) &= \int \prod_{\Rv,i} \f{\beta d \lambda_{\Rv,i}}{\pi}  e^{- i \beta \lambda_{\Rv,i} \left( \Sv_{\Rv,i} \cdot \Sv_{\Rv,i} -1 \right)},
\end{align}
where we for convenience have scaled the integration variables by the inverse temperature $\beta$.
Then defining the Fourier-transformed constraint field $\lambda_{\qv,i}$ so that
\be
\lambda_{\Rv,i} = \sum_{\qv} \lambda_{\qv,i} e^{i \qv \cdot \Rv}.
\ee
The sum in the exponent of the integrand can then be written in terms of Fourier transformed quantities
\be
\sum_{\Rv,i} \lambda_{\Rv,i} \left( \Sv_{\Rv,i} \cdot \Sv_{\Rv,i} -1 \right) =
\sum_{\qv \neq 0, i} \lambda_{\qv,i}  \left( \sum_{\qvp} \Sv^*_{\qv+\qvp,i} \cdot \Sv_{\qvp,i} \right)
+ \sum_i \lambda_{\qv=0,i} \left( \sum_{\qvp} \Sv^*_{\qvp,i} \cdot \Sv_{\qvp,i} -N_c \right),
\ee
where the $\qv=0$ components have been written separately.
The quantity $\sum_{\qvp} \Sv^*_{\qv+\qvp,i} \cdot \Sv_{\qvp,i}$ can be interpreted as the spatial modulation of the squared spin length on sublattice $i$ with wave vector $\qv$. Thus the integrations over $\lambda_{\qv \neq 0,i}$ force these modulations to be zero, i.e. no spatial variations of the (squared) spin lengths. In contrast the $\lambda_{\qv=0,i}$ integration forces the sum of the squared spin lengths on sublattice $i$ to add up to $N_c$. Together they thus enforce the local constraint that each spin has unit length. We will treat the integrations over $\lambda_{\qv \neq 0,i}$ in an approximate way using diagrams, and the integrations over $\lambda_{\qv=0,i}$ using the saddle-point method. To emphasize this distinction, we will use another symbol for the $\qv=0$ components: $\lambda_{\qv=0,i} \equiv -i \Delta_i$, and from now on interpret $\lambda_{\qv,i}$ as having zero $\qv=0$ components. Putting everything together, the partition function becomes
\be
Z = \int D \Delta D\lambda DS   \; e^{-S}
\ee
with
\be
S = \sum_{\qv,\qvp,\alpha,i,j} S^{\alpha *}_{\qv,i} \left[ \left( J_{\qv,ij} +  \Delta_i \delta_{ij} \right) \delta_{\qv,\qvp} -(-i) \lambda_{\qv-\qvp,i} \delta_{ij}  \right] S^\alpha_{\qvp,j} 
-\beta N_c \sum_i \Delta_i, \label{thisaction}  
\ee
where the spins have been rescaled by a factor $\sqrt{\beta}$.

In order to construct a diagrammatic theory, we introduce the bare inverse spin propagator $\Kbaremat$, which is a matrix in the combined $\qv$ and sublattice space with matrix elements $\Kbare_{(\qv i) (\qvp j)} = \Kbare_{\qv,ij} \delta_{\qv,\qvp}$, where
\be
\Kbare_{\qv,ij} \equiv J_{\qv,ij} + \Delta_{i} \delta_{ij}.
\ee
We also define the constraint field $\Lambdabaremat$ as a matrix containing the constraint field as components, with matrix elements
\be
    \Lambdabare_{(\qv i) (\qvp j)} = -i \lambda_{\qv-\qvp, i} \delta_{ij}.
\ee
Generalizing the number of spin components to $N_s$ and integrating over the spins, we arrive at the following expression for the partition function
\be
Z = \int  D\Delta D\lambda \; e^{-\left( S_0 + S_2 + S_r \right)},
\ee
where we have omitted field-independent constants, and divided the remainder into terms according to their powers of $\Lambdabaremat$ so that
\begin{align}
S_0 & =-\beta N_c  \sum_i \Delta_i + \f{N_s}{2} \Trace \ln \Kbaremat, \\
S_2 &= - \f{N_s}{2 \cdot 2} \Trace \left( \Kbareinvmat \Lambdabaremat \Kbareinvmat \Lambdabaremat \right), \\
S_r &= -\sum_{l=3}^\infty \f{N_s}{2 \cdot l} \Trace \left( \Kbareinvmat \Lambdabaremat \right)^l.
\end{align}
The $\Trace$-symbol indicates the trace over the combined $\qv$ and sublattice space.
There is no $S_1$-term as the constraint field has no $\qv=0$ components.
The term $S_2$ can be written
\be
S_2 = \f{N_s}{2 \cdot 2} \sum_{\qv,i} \sum_{\qvp,j} \Kbareinv_{\qv,ij} \lambda_{\qv-\qvp,j} \Kbareinv_{\qvp,ji} \lambda_{\qvp-\qv,i}
= \f{N_s}{2 \cdot 2} \sum_{\qv,i} \sum_{\qvp,j} \lambda_{\qvp-\qv,i} \Kbareinv_{\qv,ij}  \Kbareinv_{\qvp,ji} \lambda_{\qv-\qvp,j}.
\ee
Changing summation variables $\qv \to \qv + \qvp$ and $\qvp \to \pv$, one gets
\be
S_2 = \f{N_s}{2 \cdot 2} \sum_{\qv,i} \sum_{\pv,j} \lambda_{-\qv,i} \Kbareinv_{\qv+\pv,ij}  \Kbareinv_{\pv,ji} \lambda_{\qv,j}, 
\ee
and $S_2$ defines therefore the bare inverse constraint-field propagator $\Dbareinvmat$ with components $\Dbareinv_{(\qv i) (\qvp j)} = \Dbareinv_{\qv ij} \delta_{\qv,\qvp}$, where
\be
   \Dbareinv_{\qv ij} = \f{N_s}{2} \sum_{\pv} \Kbareinv_{\pv+\qv,ij}  \Kbareinv_{\pv,ji}.
\ee
Diagrammatically we will represent the bare constraint-field propagator $\Dbare_{\qv, ij}$ as a thin wavy line, and the spin propagator $\Kbareinv_{\qv, ij}$ as a thin solid line with an arrow, carrying momentum $\qv$ (in the arrow direction) from $i$ to $j$.
Diagramatically the term $S_r$ for $r \geq 3$ is a ring of thin solid line segments having $r$ hooks to attach wavy lines to. Such a ring carries a factor $N_s$ and a wavy line carries a factor $1/N_s$.   

\begin{figure}[]
\includegraphics[width=0.5\columnwidth]{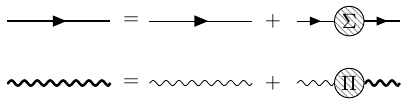}
\caption{Dyson equations for the renormalized spin propagator $K^{-1}_{\qv, ij}$
(bold solid line), and the renormalized constraint-field propagator $D_{\qv, ij}$ (bold wavy line). \label{fig:Dyson}}
\end{figure}

\begin{figure}[]
\includegraphics[width=0.36\columnwidth]{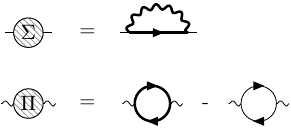}
\caption{Self-consistent equations for the self-energy $\Sigma_{\qv,ij}$ and polarization $\Pi_{\qv,ij}$. \label{fig:selfconsistentdiagrams}}
\end{figure}

In order to capture symmetry-breaking phenomena going beyond simple perturbation theory, we construct a diagrammatic theory with renormalized propagators.
In particular, $\Kinv_{\qv,ij}$ refers to the renormalized spin propagator with a self-energy addition $\Sigma_{\qv,ij}$ in its denominator 
\be
K_{\qv,ij} = J_{\qv,ij} + \Delta_i \delta_{ij}  +\Sigma_{\qv,ij} \label{Keq}.
\ee
Note that in comparison to Ref.~\onlinecite{Glittum2023} we have changed the sign on the definition of the self-energy.
The renormalized constraint-field propagator is written in terms of the renormalized spin propagators as
\be
\Dinv_{\qv,ij} = \f{N_s}{2} \sum_{\pv} \Kinv_{\qv+\pv,ij} \Kinv_{\pv,ji}, \label{Deq}
\ee
and the self-energy is taken to be the dressed Fock diagram
\be
\Sigma_{\qv,ij} = \sum_{\pv \neq 0} \Kinv_{\qv-\pv,ij} D_{\pv,ij}, \label{Sigmaeq}
\ee
calculated with fully dressed propagators, but leaving out vertex corrections. 
Diagramatically these renormalizations correspond to the resummations shown in Fig.~\ref{fig:Dyson}, with the self-energy and polarization shown in Fig.~\ref{fig:selfconsistentdiagrams}.

\begin{figure}[t]
\includegraphics[width=0.4\columnwidth]{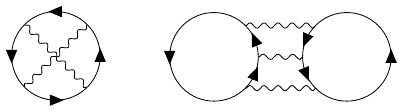}
\caption{Leading-order omitted diagrams. Wavy lines indicate the constraint-field propagator $\Dbare$ which contains a factor $1/N_s$. Solid lines indicate the spin propagator $\Kbareinv$. A closed solid loop carries a factor $N_s$. \label{DiagramsSR}}
\end{figure}

The constraint field can then be formally integrated out, which gives
\be
Z = \int D \Delta \; e^{-S^\prime},
\ee
where
\be
S^\prime = -\beta N_c  \sum_i \Delta_i + \f{N_s}{2} \sum_{\qv} \ln{ \det K_{\qv}}
+ \f{1}{2} \sum_{\qv \neq 0} \ln{\det{D^{-1}_{\qv}}}-\f{N_s}{2} \sum_{\qv} \trace \left( K^{-1}_{\qv} \Sigma_{\qv} \right). \label{Sprime}
\ee
We have here employed the notation where $K_{\qv}$ means a three-by-three sublattice matrix with matrix elements $K_{\qv,ij}$ (similar for $D^{-1}_{\qv}$ and $\Sigma_{\qv}$), and the $\det$($\trace$)-symbol indicates taking the determinant(trace) in this sublattice space.
To arrive at this expression, we have neglected diagrams in a systematic large-$N_s$ expansion as described in appendix A of Ref.~\onlinecite{Glittum2021}. The leading-order diagrams being omitted (shown in Fig.~\ref{DiagramsSR}), and hence the error in the free energy, are of order $1/N_s$.

The final integrals over the $\Delta_i$s are calculated using the saddle-point method. Differentiating $S^\prime$ w.r.t. $\Delta_i$ we arrive at the three saddle-point equations giving the inverse temperature
\be
    \beta = \f{N_s}{2N_c} \sum_{\qv} \Kinv_{\qv, ii}, \label{betaeq}
\ee
where no sum over $i$ is implied. This simple form is a consequence of the fact that the contributions from the two last terms in Eq.~(\ref{Sprime}) cancel each other (cf. Ref.~\onlinecite{Glittum2021}).
The three equations Eq.~(\ref{betaeq}) enforce the average spin length on each sublattice $i$ to be unity, and they must all give the same temperature for the solution to be physical. It should be noted that since $\Sigma_{\qv}$ is not fixed, it is possible to have several sets $\{ (\Sigma_{\qv}^{(1)},\Delta^{(1)}),(\Sigma_{\qv}^{(2)},\Delta^{(2)}),\ldots\}$ which give the same value for the sum in Eq.~(\ref{betaeq}), i.e. the same (inverse) temperature. Given this possible multivaluedness, we view Eq.~(\ref{betaeq}) as equations that give the temperature for a given value of the $\Delta_i$s and $\Sigma_{\qv}$.

The self-consistent equations \eqref{Keq}, \eqref{Deq} and \eqref{Sigmaeq} are solved by iteration, starting typically from a random self-energy and equal values of the $\Delta_i$s. Each iteration gives an overall positive contribution to the self-energy. To avoid the general increase in temperature associated with this, the $\Delta_i$s are renormalized in every iteration by subtracting from them the minimum eigenvalue among all $\Sigma_{\qv}$ matrices. In addition, each $\Delta_i$ is adjusted very slightly so that Eq.~\eqref{betaeq} give the same value of the temperature for all sublattices. We iterate until the temperature in subsequent iterations has converged, and then employ the converged 
$\Kinv_{\qv}$, $\Sigma_{\qv}$ and $D_{\qv}$ to calculate the free energy. Specifically we use the convergence criterion that three subsequent iterations are required to have temperatures that differ by at most by $10^{-13}$. Typically, $10-200$ iterations are needed for convergence. 

After reaching convergence, we calculate the free energy per spin at the converged temperature $T$ as follows
\be \label{eq: f}
f = - \f{N_sT}{2N} \sum_{\qv} \ln\det\left( \pi T \Kinv_{\qv}\right) + \f{T}{2N}\sum_{\qv}\ln\det\left(\f{\pi T^2}{2N_c} D^{-1}_{\qv}\right) -\f{N_c}{N}\sum_{i} \Delta_i
- \f{N_s T}{2N} \trace \left(\Kinv_{\qv}\; \Sigma_{\qv}\right),
\ee
where the renormalized values of the $\Delta_i$s should be used. We have also reinstated factors from the integration measures.

To get information about the spin correlations we calculate the quantity
\be
A_{\qv} \equiv \sum_{ij} \Kinv_{\qv, ij} e^{-i \qv \cdot (\vec{\alpha}_i -\vec{\alpha}_j)},
\ee
which is closely related to the spin structure factor $S(\qv) \equiv \sum_{ij} \langle \Sv_{-\qv,i} \cdot \Sv_{\qv,j} \rangle e^{i \qv \left( \alpha_i -\alpha_j \right)}  =N_s T \left(A_{\qv} + A_{-\qv} \right)/4$.


\subsection{Spin-liquid real-space correlations}

\begin{figure}[]
\begin{center}
\includegraphics[height=0.45\columnwidth]{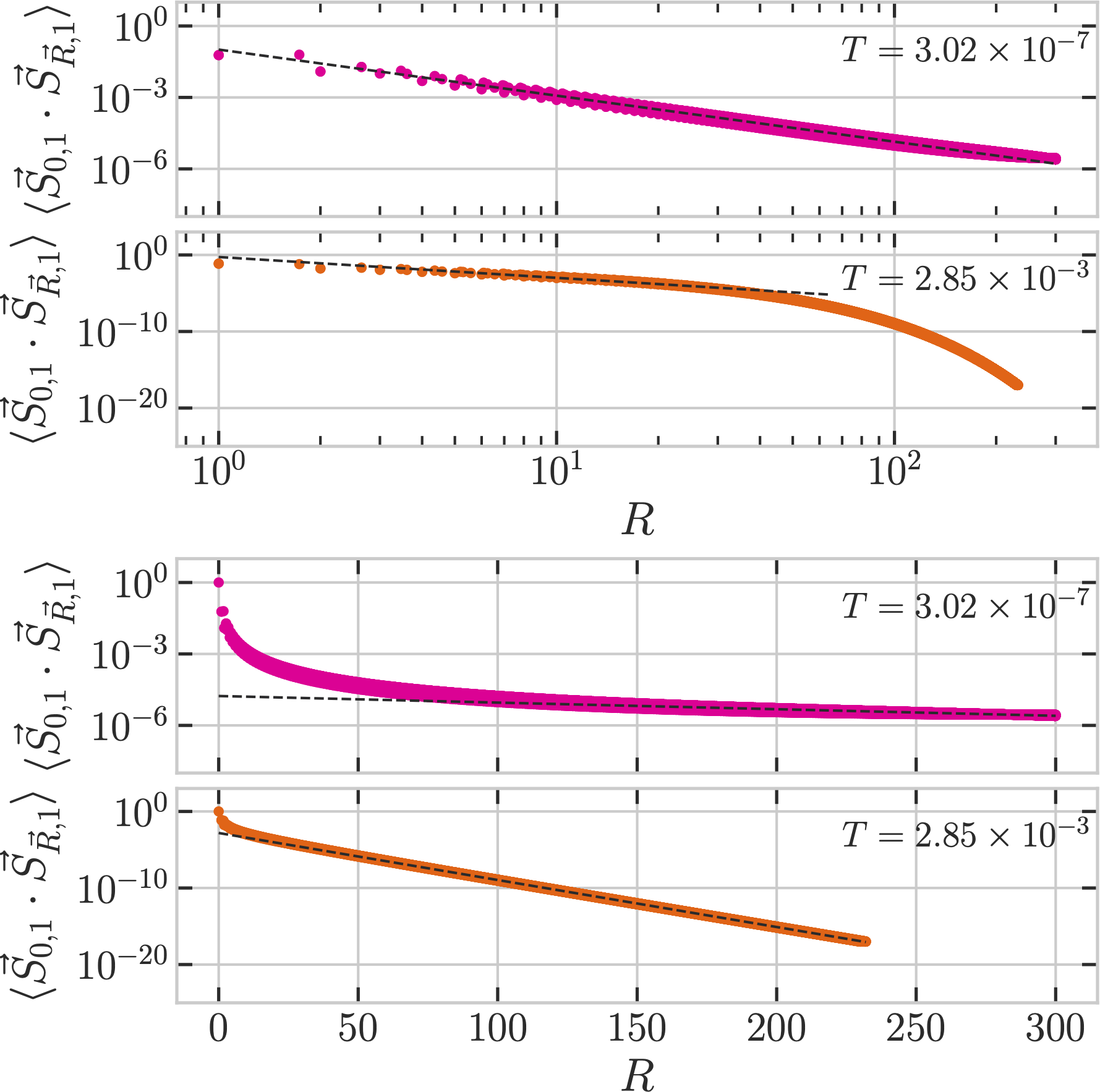}
\includegraphics[height=0.45\columnwidth]{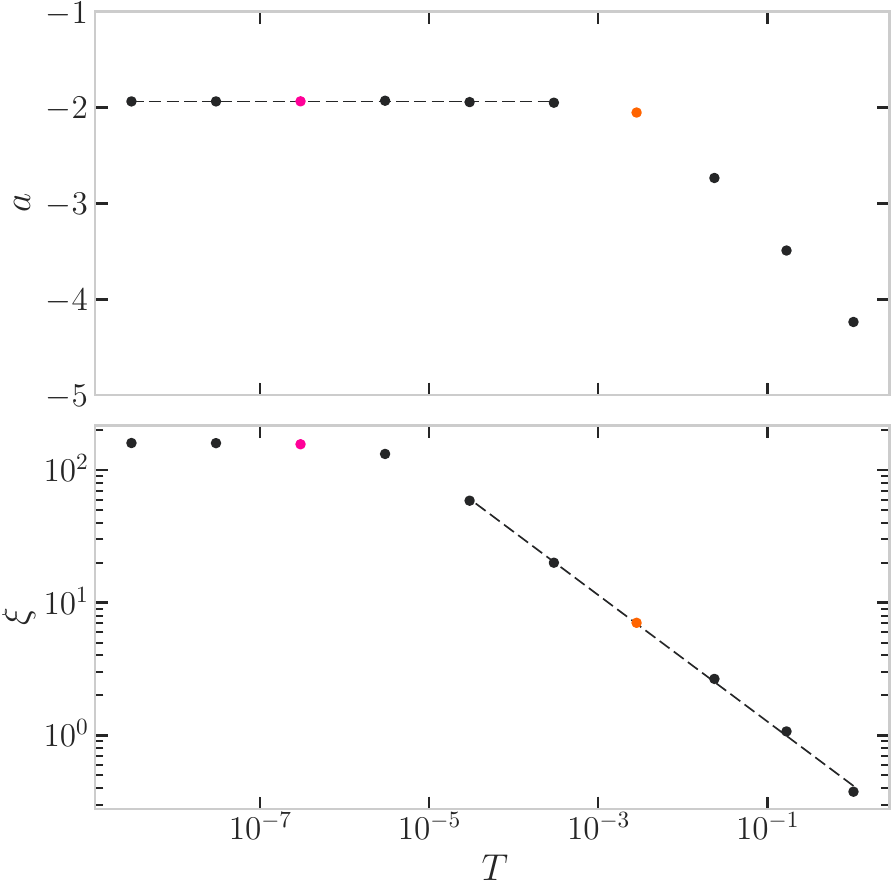}
\caption{Left: Real-space correlations in the spin liquid at sublattice 1 as function of distance $R$ between the spins for two different temperatures. Values smaller than $10^{-17}$ are not shown. Right: Extracted values for the power-law exponent $a$ and correlation length $\xi$ as function of temperature.
$L=600$.
}
\label{fig:realspacecorrelations}
\end{center}
\end{figure}

We compute the real-space correlations in the spin liquid as
\begin{equation}
\langle \Sv_{0,i} \cdot \Sv_{\Rv, j} \rangle = \frac{1}{N_c}\sum_{\qv} \langle \Sv_{-\qv,i} \cdot \Sv_{\qv, j}\rangle e^{i\qv\cdot \Rv} = \frac{N_sT}{4N_c} \sum_{\qv} (\Kinv_{-\qv, ij} + \Kinv_{\qv, ji}) e^{i \qv \cdot \Rv},
\end{equation}
and fit them to the functional form 
\begin{equation}
\langle \Sv_{0,i} \cdot \Sv_{\Rv, j} \rangle \propto \frac{1}{R^a}e^{-R/\xi},
\end{equation}
where $R \equiv |\Rv|$, and $a$ is an exponent characterizing a power-law decay at distances smaller than the correlation length $\xi$.

For low temperatures, the correlation length is expected to be long, and consequently $\langle \Sv_{0,i} \cdot \Sv_{\Rv, j} \rangle \propto 1/R^a$. $a$ can thus be extracted as the slope in a log-log plot, see Fig.~\ref{fig:realspacecorrelations} left upper panel. We find that the straight-line fits work well for $T < 10^{-3}$. The resulting exponents are shown in the right upper panel, giving $a = 1.94 \approx 2$, in accordance with Ref.~\onlinecite{Garanin1999}. For higher temperatures, the correlation length is small, and the exponential decay dominates. For these temperatures, we can find the correlation length by extracting the negative inverse of the slope in a semi-log plot, see left lower panel. These fits work well for $T> 10^{-5}$. The resulting correlation lengths for different temperatures are shown in the right lower panel as a log-log plot.  We find a slope of $-0.48 \approx -1/2$. Thus, the correlation length is estimated to go as $\xi \propto 1/\sqrt{T}$, also in accordance with Ref.~\onlinecite{Garanin1999}.

\subsection{Finite-size behavior of the critical temperature and latent heat}

\begin{figure}[]
\begin{center}
\includegraphics[width=0.45\columnwidth]{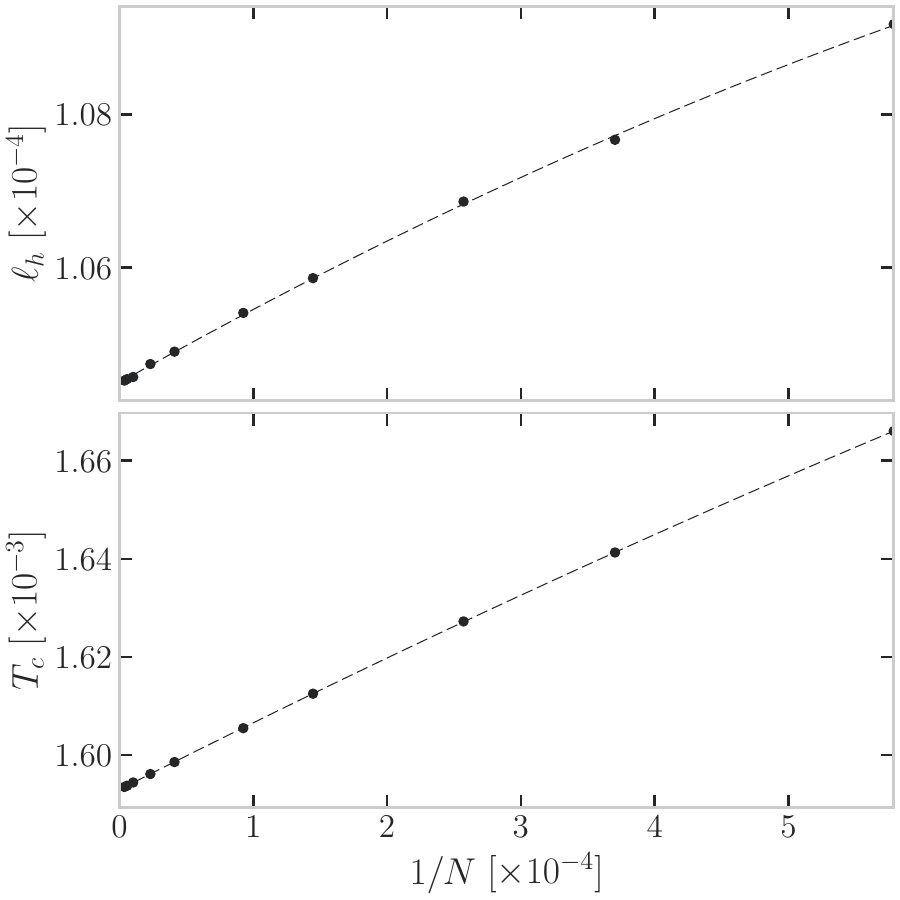}
\caption{Upper panel: The latent heat per spin of the phase transition vs. inverse system size.
Lower panel: $T_c$ vs. inverse system size. The quadratic fits extrapolate to a latent heat of $1.04499 \cdot 10^{-4}$ and a critical temperature $T_c= 1.59294 \cdot 10^{-3}$ for infinite system size.}
\label{fig:latentheat}
\end{center}
\end{figure}

The critical temperature $T_c$ and latent heat per spin $\latentheat$ are extracted from the crossing of the spin liquid and the \rthree free-energy branches. $T_c$ is determined as the crossing temperature, and the difference in slopes of the two branches at the crossing is multiplied by $T_c$ to give the latent heat per spin. The results for different system sizes are plotted vs. the inverse system size $1/N$ in Fig.~\ref{fig:latentheat}. The points are fitted to a quadratic polynomial in $1/N$ (shown as the lines in the plots) and the infinite-size behavior is extracted, see figure caption.

\subsection{Low-temperature behavior of the specific heat}

\begin{figure}[]
\begin{center}
\includegraphics[width=0.45\columnwidth]{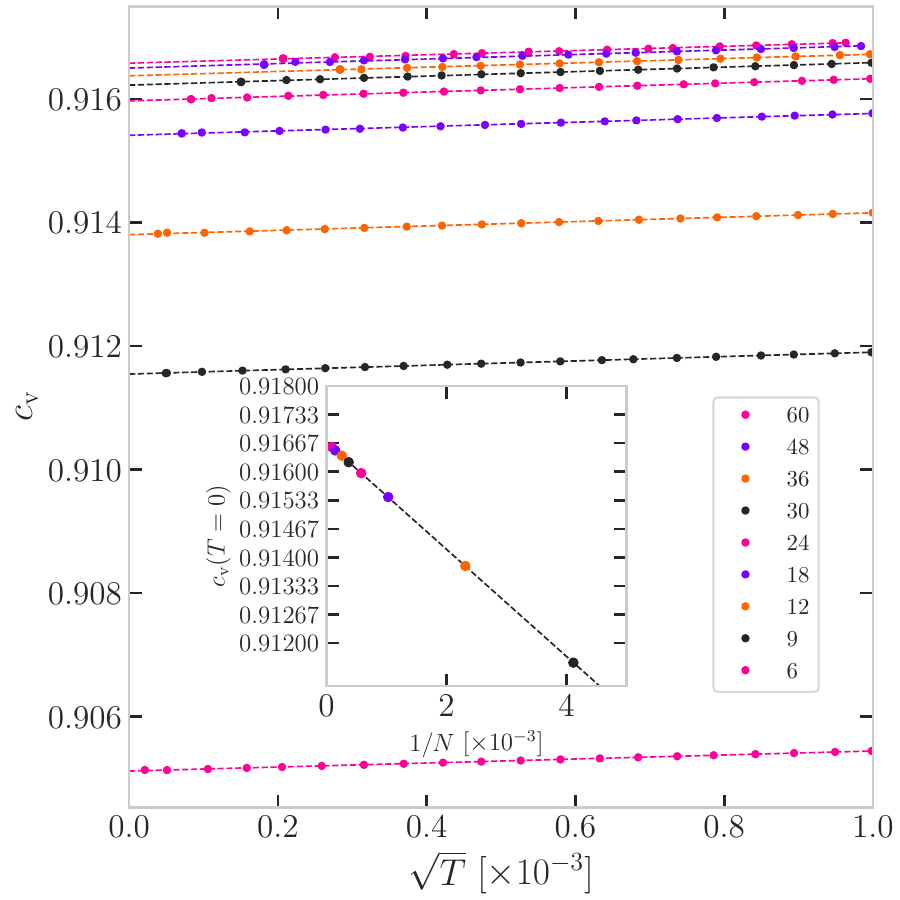}
\caption{Specific heat vs. square root of temperature at low temperatures for different linear system sizes $L$ indicated by the legends. The dashed lines show fits to the functions $\cv (T)=a+b \sqrt{T}$. The inset (also shown in the main text in Fig.~\ref{fig:specificheat} upper inset) shows the $y$-intercepts $a$ vs $1/N$ $(N=3L^2)$, and the dashed line is a linear fit to these points giving $\cv (T = 0) = 0.9167 - 1.2495/N$.}
\label{fig:finitesizecvlow}
\end{center}
\end{figure}

In Fig.~\ref{fig:finitesizecvlow}, we show the specific heat $\cv$ in the \rthree phase at very low temperatures ($T<10^{-6}$) for several system sizes indicated by the legends.  We have plotted $\cv$ vs. $\sqrt{T}$ and fitted the points to straight lines which works very well at these low temperatures. The $y$-axis intercepts of the fitted lines are plotted in the inset vs. inverse system size and fitted to a straight line $\cv (T = 0) = 0.9167 - 1.2495/N$.

\subsection{Crossover peak for $J_2=-0.02$}

\begin{figure}[]
\begin{center}
\includegraphics[height=0.45\columnwidth]{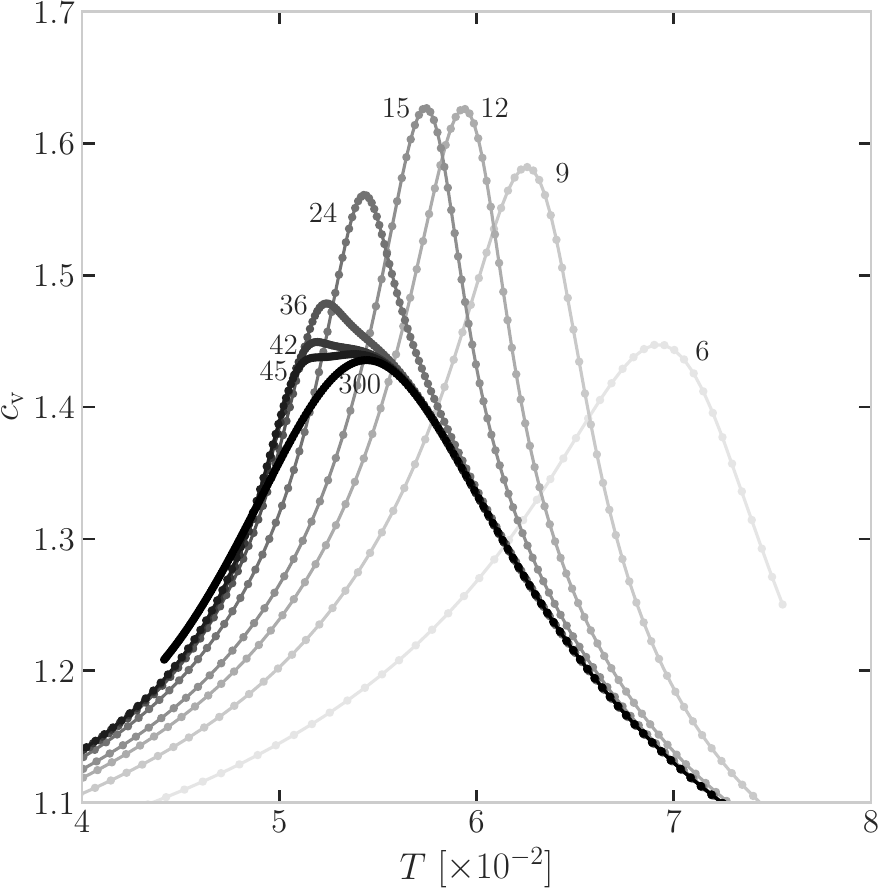}
\includegraphics[height=0.45\columnwidth]{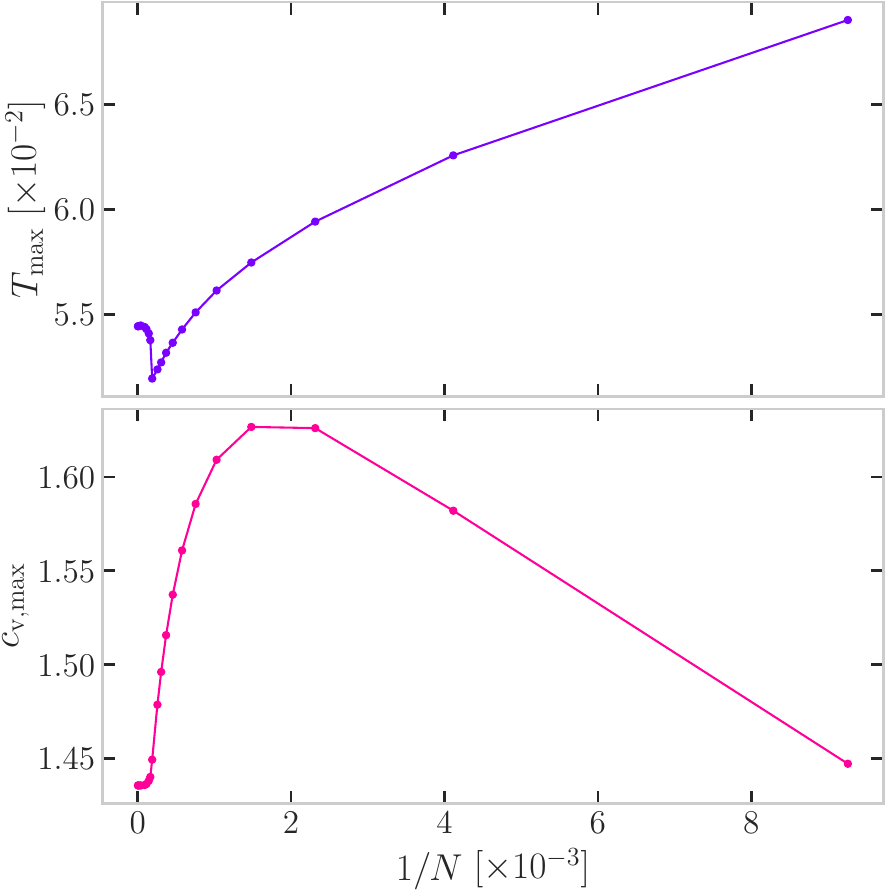}
\caption{
Left: Specific heat peaks at $J_2=-0.02$ for different linear system sizes $L$ indicated by the numbers and gray scale (darker curves correspond to larger system sizes). 
Right: Temperature and height of the corresponding peak maxima vs. inverse system size.
}
\label{fig:crossoverpeak}
\end{center}
\end{figure}

For $J_2$ values lying outside the critical points, the specific heat exhibits a broad peak at the dashed lines in Fig.~\ref{fig:phasediagram} in the main text. For $J_2=-0.02$ this peak was studied in Ref.~\onlinecite{Spenke2018} for linear system sizes $L\leq 24$. They concluded that the peak indicates a phase transition in the 2D Ising universality class. We show in Fig.~\ref{fig:crossoverpeak} that NBT on the contrary reveals substantial finite-size effects in this system which should not be interpreted as indications of a phase transition.  Fig.~\ref{fig:crossoverpeak} (left) shows the specific heat peak obtained in NBT vs. temperature at $J_2=-0.02$ for different linear system sizes. As in Ref.~\onlinecite{Spenke2018}, for small $L \leq 15$, the peak grows and moves down in temperature as $L$ increases. However, when increasing $L$ further, we find that the peak shrinks, and for $L \simeq 36-45$ its shape reveals two features; a low-temperature peak that shrinks and becomes a shoulder before it eventually disappears for larger system sizes, and a robust feature at a slightly higher temperature that develops into a stable broad peak that do not evolve further with system size. We take this latter broad peak to indicate a crossover rather than a phase transition.
In the right panel of Fig.~\ref{fig:crossoverpeak} we summarize how the temperature and height of the peak maximum behave with inverse system size. The marked change around $L=42$ is a consequence of the development of the stable broad peak.

\subsection{Pyrochlore antiferromagnet}

\begin{figure}[]
\begin{center}
\includegraphics[width=0.4555\columnwidth]{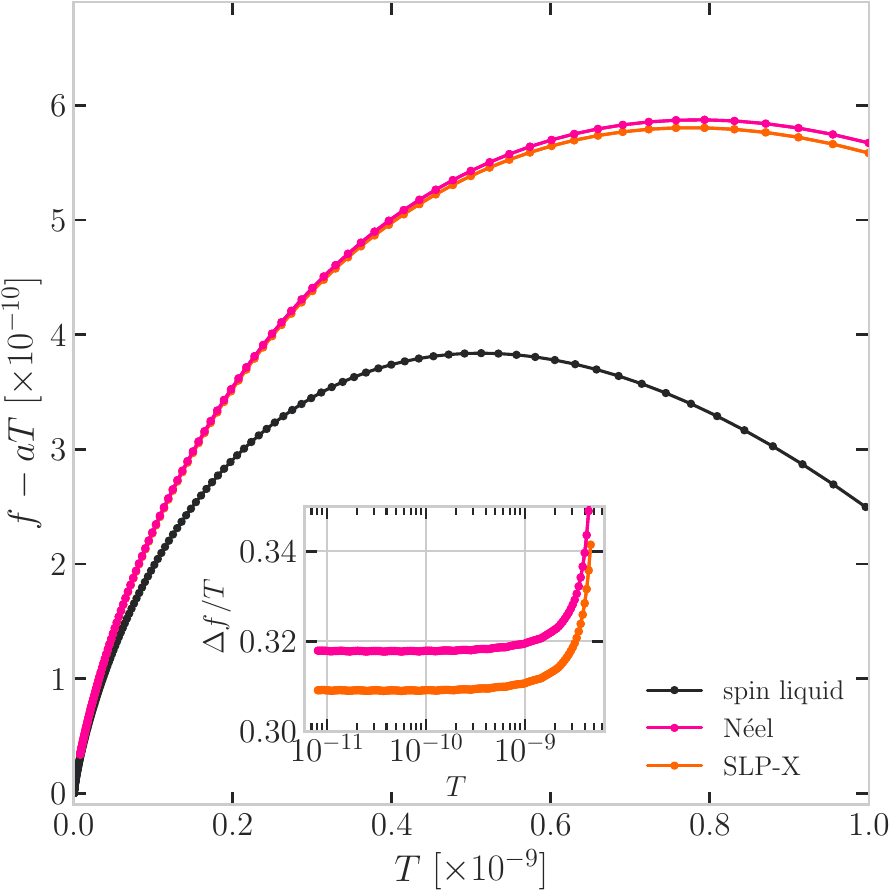}
\includegraphics[width=0.45\columnwidth]{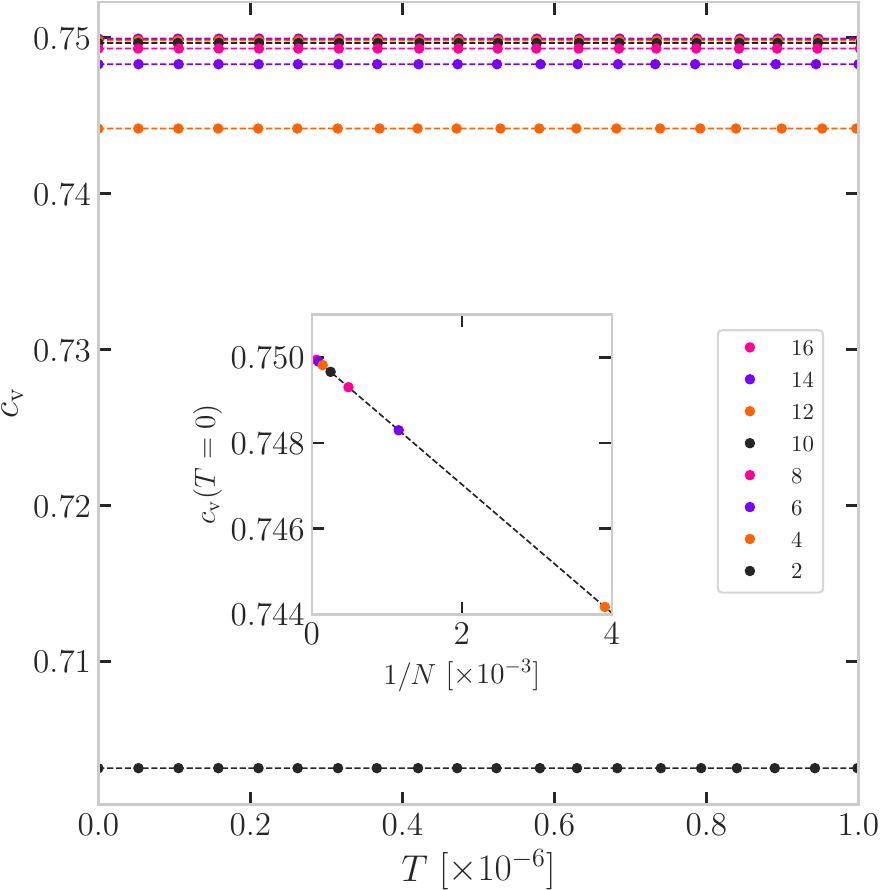}
\caption{Left: Free energy per spin for the pyrochlore antiferromagnet for the different phases indicated by the legends. The inset shows the free-energy difference of the biased curves from the spin liquid and is divided by temperature. A term $aT$ with $a=13$ has been subtracted in the main plot to better visualize the difference between the free energies. $L=12$.
Right: Pyrochlore specific heat vs. temperature at low temperatures for different linear system sizes $L$ indicated by the legends. The dashed lines show fits to the functions $\cv (T)=a+b T$. Although difficult to see, the values of $b$ are approximately $1$. The inset shows the $y$-intercepts $a$ vs $1/N$ $(N=4L^3)$, and the dashed line is a linear fit to these points giving $\cv (T=0) = 0.7500 - 1.5001/N$.}
\label{fig:pyrofreeenergy}
\end{center}
\end{figure}

We have also used NBT for the pyrochlore nearest-neighbor antiferromagnet and compared free energies of the spin liquid with the Néel~\cite{Chern2008,Lapa2012,Iqbal2019} and SLP-X~\cite{Ghosh2019, Glittum2023} states, which are colinear ground states in the vicinity of the pure antiferromagnetic point, see left panel of Fig.~\ref{fig:pyrofreeenergy}. For temperatures $T > 10^{-10}$, it is clear that the free energies of the competing states is higher than for the spin liquid. In the inset, we show $\Delta f/T$, the free-energy difference between the competing states and the spin liquid divided by temperature, which stays almost constant at a positive value as $T \to 0$. We cannot rule out that these lines will eventually cross zero, but based on the flatness of the curves, it can only happen at extremely low temperatures.

We have also repeated the specific heat calculation for the pyrochlore lattice, see right panel of Fig.~\ref{fig:pyrofreeenergy}, which shows the specific heat for the spin liquid at very low temperatures for different system sizes. In contrast to the kagome lattice, $\cv$ behaves linearly on $T$ at low temperatures. The extrapolated $T = 0$ values of the specific heat are plotted in the inset vs. inverse system size and fitted to a straight line, $\cv (T=0) = 0.7500 - 1.5001/N$.

\subsection{Search for alternative stable states}

\begin{figure}[]
\begin{center}
\includegraphics[width=0.45\columnwidth]{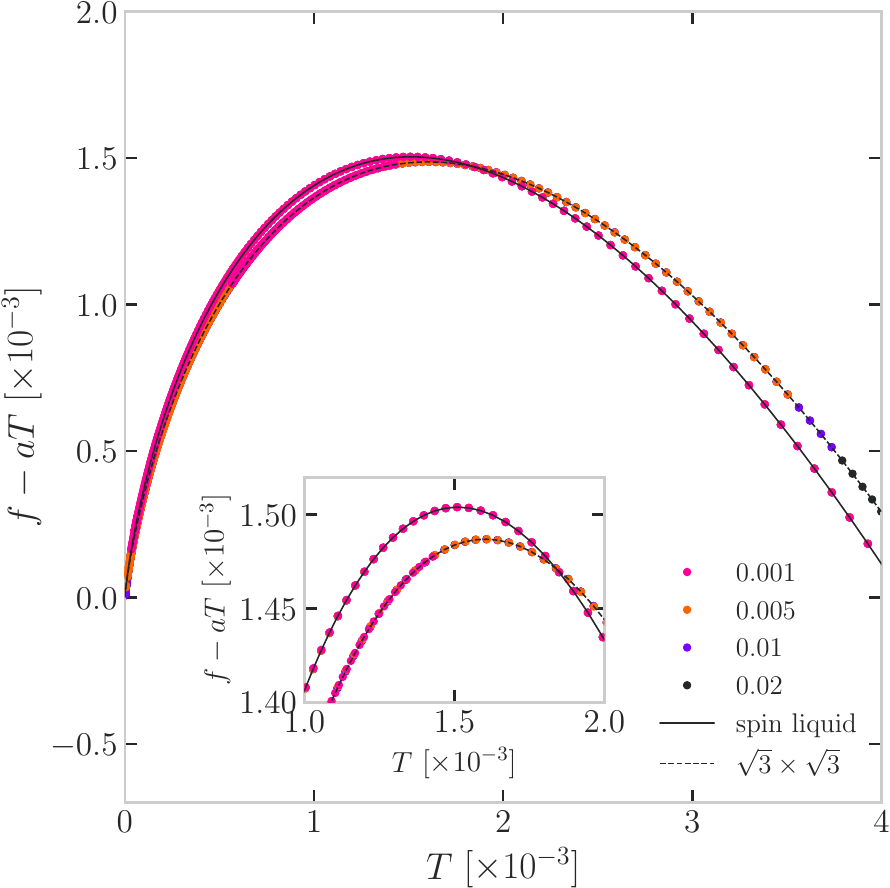}
\caption{Free energy per spin for NBT simulations initiated with self-energies gotten from MC simulations at temperatures indicated by the legends (colored circles). Here the average MC correlation function is used. 
The solid and dashed lines are the free energies obtained by NBT of the spin liquid and \rthree phase, respectively. $L=12$. $a=3$. The inset shows a blow-up of the phase-transition region.}
\label{fig:mcbias}
\end{center}
\end{figure}

NBT relies on convergence from an initial self-energy. This has the advantage that free energies of different states can be obtained -- also those that are metastable. However, a disadvantage is that one cannot know that all states with low free energy have been found. As described in the main text, we have searched for low free-energy states by running NBT for slightly perturbed systems, and picked the converged self-energy of those as initial self-energies for the nearest-neighbor model at low values of $\Delta$. This gives the free energies shown in Fig.~\ref{fig:freeenergy} in the main text.

We have also biased NBT with initial self-energies obtained from Monte Carlo (MC) simulations.
The MC simulations were carried out starting from a random spin state at high temperature, and then gradually lowering the temperature to the desired one. We used four different temperatures $T= \{0.001,0.005,0.01,0.02\}$, and a linear system size $L=12$.  
We used the Metropolis algorithm with updates in which the proposed spin is selected in a cone around the old spin~\cite{Hinzke1999,LandauBinder2009}. The cone is adjusted during equilibration to get acceptance probability one-half~\cite{Alzate-Cardona2019}. After equilibrating for $10^8$ MC steps, the spin correlation function was calculated and converted to a self-energy. We used correlation functions based on both single-configuration snapshots, and on MC averages (for another $10^8$ MC steps). In all, three initial self-energies (two based on independent MC snapshots, and one based on MC averages) were investigated for each temperature. 

For each of the twelve different MC initial self-energies, we ran 800 NBT simulations with different $\Delta$-values equally spaced on a log-scale in the interval $[10^{-9},1]$, all starting with the same initial self-energy.
The converged free energies and temperatures were recorded. They are shown as colored circles in Fig.~\ref{fig:mcbias} for the case where the initial self-energies are based on the average over $10^8$ MC configurations. The runs initiated with MC snapshots are similar. We find that {\em all} these runs either converge to the spin-liquid branch or to the \rthree branch. For the runs initiated by MC snapshots, there is a trend that the NBT initiated by low-temperature configurations has a tendency of converging to the \rthree branch. However, and perhaps somewhat surprisingly, the temperature of the MC configurations do not seem to play any significant role in which state the NBT converges to when the inital self-energy is based on MC {\em averages}, at least not for the four temperatures we have investigated.\\ \newline

\putbib
\end{bibunit}

\end{document}